\documentclass[prl,twocolumn,showpacs,superscriptaddress,amsfonts,amsmath,floatfix]{revtex4-1}

\usepackage{graphicx}
\usepackage{bbold}
\usepackage{color}

\begin{document}


\title{Weyl points in three-dimensional optical lattices:\\ synthetic magnetic monopoles in momentum space}

\author{Tena Dub\v{c}ek}
\affiliation{Department of Physics, University of Zagreb, Bijeni\v{c}ka c. 32, 10000 Zagreb, Croatia}
\author{Colin J. Kennedy}
\affiliation{Department of Physics, Massachusetts Institute of Technology, Cambridge, Massachusetts 02139, USA}
\author{Ling Lu}
\affiliation{Department of Physics, Massachusetts Institute of Technology, Cambridge, Massachusetts 02139, USA}
\author{Wolfgang Ketterle}
\affiliation{Department of Physics, Massachusetts Institute of Technology, Cambridge, Massachusetts 02139, USA}
\author{Marin Solja\v{c}i\'{c}}
\affiliation{Department of Physics, Massachusetts Institute of Technology, Cambridge, Massachusetts 02139, USA}
\author{Hrvoje Buljan}
\affiliation{Department of Physics, University of Zagreb, Bijeni\v{c}ka c. 32, 10000 Zagreb, Croatia}

\date{\today}

\begin{abstract}
We show that Hamiltonians with Weyl points can be realized for ultracold atoms using 
laser-assisted tunneling in three-dimensional optical lattices.
Weyl points are synthetic magnetic monopoles that exhibit a robust, three-dimensional 
linear dispersion. 
They are associated with many interesting topological states of matter, such as Weyl 
semimetals and chiral Weyl fermions.
However, Weyl points have yet to be experimentally observed in any system. 
We show that this elusive goal is well-within experimental reach with an extension 
of the techniques recently used to obtain the Harper Hamiltonian.
\end{abstract}

\maketitle

The Weyl Hamiltonian was introduced in 1929~\cite{Weyl1929} in particle physics 
as a simplification of the relativistic Dirac equation~\cite{Dirac1928} to 
describe massless chiral fermions. 
It was conjectured that neutrinos could be Weyl fermions before the discovery 
of neutrino oscillations ruled out such a possibility.
Interest in massless chiral fermions continued in the condensed matter literature in the 
form of an effective Hamiltonian describing the band structure of theoretically predicted topological 
Weyl semimetals in proximity of gap closing points, also referred to as Weyl points, 
in the Brillouin zone~\cite{Wan2011, AshwinReview, HosurReview}. 
These systems followed the development of topological insulators~\cite{Kane2005, HasanReview}, 
emphasizing the role of band topology in describing exotic phases of matter. 
Important manifestations of these phases are surface states, which in the case of Weyl semimetals 
take the intriguing form of 'Fermi arcs'~\cite{AshwinReview, HosurReview}. 
Weyl semimetals were predicted in pyrochlore iridates~\cite{Wan2011,AshwinReview} and 
heterostructures based on topological insulators~\cite{AshwinReview}. 
However, as a consequence of the complicated structure and symmetry requirements of 
the candidate materials, despite a very large body of theoretical interest, 
Weyl semimetals and massless chiral Weyl fermions have not 
yet been observed in nature.

Recent experiments on synthetic magnetic/gauge fields in ultracold atomic 
gases~\cite{Madison2000, Abo2001, Lin2009, Aidelsburger2011, Struck2012, Miyake2013, Aidelsburger2013, Aidelsburger2014, Jotzu2014, Ray2014}, 
alongside advances in photonics~\cite{Wang2009, Fang2012, Zeuner2012, Kraus2012, Rechtsman2013, Hafezi2013, Lu2013}, 
have propelled these systems as promising platforms for investigating topological effects and novel states of matter 
(see Refs.~\cite{Dal2011, Bloch2012, Goldman2014, Carusotto2013, Lu2014} for reviews). 
However, Weyl points have been scarcely addressed in these fields, and only 
theoretically~\cite{Lu2013, Lan2011, Anderson2012, Jiang2012, Ganeshan2014}. 
In photonics, a double gyroid photonic crystal with broken time-reversal 
and/or parity symmetry was predicted to have Weyl points~\cite{Lu2013}. 
Theoretical lattice models possessing Weyl points~\cite{Lan2011, Jiang2012, Ganeshan2014}, 
and Weyl spin-orbit coupling~\cite{Anderson2012}, were studied in the context of 
ultracold atomic gases. Due to the elusive nature 
of Weyl semimetals and fermions and many intriguing 
novel phases and phenomena that they could enable, a viable and possibly simple 
experimental scheme for their experimental realization would be of great value. 
Here, we propose the realization of the Weyl Hamiltonian for ultracold atoms in a straightforward 
modification of the experimental system that was recently employed to obtain 
the Harper Hamiltonian~\cite{Miyake2013}.

The Harper~\cite{Harper1955} (also referred to as the Hofstadter~\cite{Hofstadter1976}) Hamiltonian 
was recently realized in optical lattices in the MIT~\cite{Miyake2013} and 
Munich~\cite{Aidelsburger2013} groups, by employing laser-assisted tunneling 
to create synthetic magnetic fields. 
Historically, the first synthetic magnetic fields were 
implemented in rapidly rotating Bose-Einstein condensates (BECs) by 
using Coriolis forces~\cite{Madison2000,Abo2001}.
The first implementation using laser-atom interactions was in the NIST group 
with spatially dependent Raman optical coupling between internal 
hyperfine atomic states in bulk BECs~\cite{Lin2009}.  
Methods of generating synthetic magnetic fields used in optical lattices engineer the complex tunneling parameters 
between lattice sites~\cite{Struck2012,Miyake2013,Aidelsburger2013}. 
They include shaking of the optical lattice, 
as demonstrated in the Hamburg group~\cite{Struck2012}, 
laser assisted tunneling which realized staggered magnetic fields
in optical superlattices~\cite{Aidelsburger2011} and the 
Harper Hamiltonian in tilted lattices~\cite{Miyake2013,Aidelsburger2013},
and an all-optical scheme which enables flux rectification in 
optical superlattices~\cite{Aidelsburger2014}.
One of the intriguing recent achievements is the realization of 
Dirac monopoles in a synthetic magnetic field produced by a bulk spinor BEC~\cite{Ray2014}. 
It should be emphasized that all lattices with nontrivial topology that were 
experimentally realized so far were in one or two dimensions. 
This work points out how a straightforward inclusion of the third dimension enables 
experiments on intriguing and elusive topological phenomena.

The laser-assisted tunneling scheme~\cite{Aidelsburger2011, Miyake2013, Aidelsburger2013} 
requires only far off-resonant lasers and a single atomic internal state, 
and thus avoids heating by spontaneous emission.
An early related proposal involved coupling of different internal states~\cite{Jaksch2003}.  
The scheme used here is based on the proposal introduced in Ref.~\cite{Kolovsky2011}, 
and later modified to enable generation of a homogeneous field~\cite{Miyake2013, Aidelsburger2013}. 
With this scheme, we can engineer both the amplitude and phase of the tunneling matrix elements in optical lattices. For example, 
if a cubic $D$-dimensional optical lattice has tunneling matrix elements $J_d$ ($d=1,\ldots,D$), 
laser assisted tunneling can in principle change them to $K_d e^{i\Phi_d}$, where the phases 
depend on the position.

\begin{figure}
\centerline{
\mbox{\includegraphics[width=0.45\textwidth]{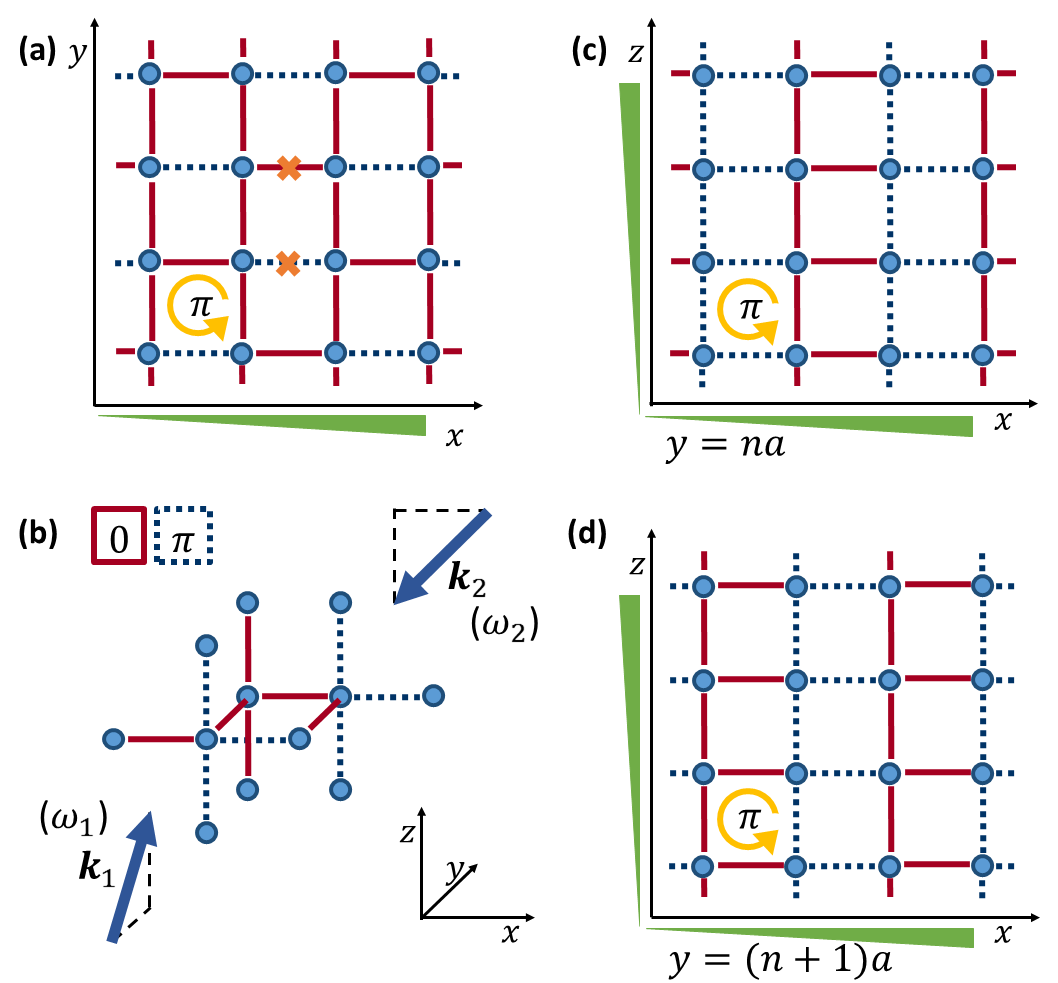}}
}
\caption{(color online) Sketch of the 3D cubic lattice with phase engineered hopping 
along $x$ and $z$ directions, which possesses Weyl points in momentum space. 
Dashed (solid) lines depict hopping with acquired phase $\pi$ ($0$, respectively). 
(a) The $xy$ planes of the lattice are equivalent to the lattice of the 
Harper Hamiltonian for $\alpha=1/2$. 
Centers of inversion symmetry for this 2D lattice are denoted by orange crosses. 
Green triangles along the axes denote the tilted directions.
(b) A pair of Raman lasers enabling laser assisted tunneling
is sketched with arrows. 
The 3D lattice can be visualized as alternating stacks of 2D lattices parallel 
to the $xz$ plane, which are shown in (c) and (d); the hopping between these 
planes (along $y$) is regular. 
The hopping along $z$ is alternating with phase $0$ or $\pi$, depending on the 
position in the $xy$ plane [see  (b), (c), and (d)], which breaks the inversion symmetry.
}
\label{lattice}
\end{figure}

For Weyl points to occur, time-reversal and/or inversion symmetry must be 
broken~\cite{AshwinReview,Lu2013}. The two-dimensional (2D) lattice realized in 
Ref.~\cite{Miyake2013}, which is our starting point, possesses both symmetries. 
Tunneling along the $x$ direction is laser assisted, 
with the phase alternating between $0$ and $\pi$, whereas 
hopping along $y$ stays regular [see Fig.~\ref{lattice}(a)]. 
The centers of inversion symmetry are denoted by orange
crosses in Fig.~\ref{lattice}(a). 
The time-reversal symmetry is a consequence of the fact that the 
accumulated phase per plaquette $\pi$ is equivalent to a phase of $-\pi$. 
This system is a realization of the Harper Hamiltonian for 
$\alpha=1/2$, where $\alpha$ is the flux per plaquette  
in units of the flux quantum~\cite{Miyake2013,Aidelsburger2013}. 
The lattice has two sublattices (A-B) giving rise to pseudospin.  
In quasimomentum representation, the Hamiltonian is  
$H_{\alpha=1/2}({\bf k}) = -2 \{ J_y \cos(k_y a)\sigma_x + K_x \sin(k_x a)\sigma_y \}$, 
where $\sigma_i$ denote Pauli matrices; it has two bands,
$E_{\alpha=1/2}=\pm 2 \sqrt{K_x^2 \sin^2(k_xa) + J_y^2 \cos^2(k_ya)}$,
touching at two 2D Dirac points at $(k_x,k_y)=(0,\pm \pi/2a)$ in the Brillouin zone~\cite{MiyakePhD2013}. 
Here $(K_x,J_y)$ denote the tunneling amplitudes, and ($k_x,k_y$) the Bloch wave vector.

Suppose that we construct a 3D lattice by stacking 2D lattices from 
Fig.~\ref{lattice}(a), one on top of each other, with regular hopping ($J_z$) along the third 
($z$) direction. This 3D lattice is described by the Hamiltonian 
\begin{equation}
H_{LN}({\bf k}) 
= -2 
\{
J_y \cos(k_y a)\sigma_x + K_x \sin(k_x a)\sigma_y  + J_z \cos(k_z a)\mathbb{1}
\},
\end{equation}
where $\mathbb{1}$ is the unity matrix. The 2D Dirac points have 
become line nodes (LN) in the 3D Brillouin zone at which the two bands touch:
$E_{LN}=- 2J_z\cos(k_za) \pm 2\sqrt{K_x^2 \sin^2(k_xa) + J_y^2 \cos^2(k_ya)}$.
Note that both the inversion and the time-reversal symmetry are 
inherited from the $\alpha=1/2$ Harper Hamiltonian. 
In order to achieve Weyl points, we must 
break one of these when adding the third dimension.

To achieve this goal, we propose to construct a 3D cubic lattice with 
laser assisted tunneling along both $x$ and $z$ directions as follows. 
First, tunneling along these directions is 
suppressed by introducing a linear tilt of energy $\Delta$ per lattice 
site, identical along $x$ and $z$. It can be obtained by 
a linear gradient potential (e.g., gravity or magnetic field gradient~\cite{Miyake2013}) 
along the $\hat x + \hat z$ direction. 
The tunneling is restored by two far-detuned Raman beams of frequency detuning 
$\delta\omega=\omega_1-\omega_2$, and momentum difference 
$\delta{\bf k}={\bf k}_1-{\bf k}_2$~\cite{Miyake2013}. For resonant tunneling, 
$\delta\omega=\Delta/\hbar$, and a sufficiently large tilt 
($J_x,J_z\ll\Delta\ll E_{gap}$)~\cite{Miyake2013}, 
time-averaging over the rapidly oscillating terms yields an effective 
3D Hamiltonian
\begin{align}
\label{Weyl H}
& H_{3D} = -\sum_{m,n,l} \left(\right. 
K_x e^{-i\Phi_{m,n,l}} a_{m+1,n,l}^{\dagger} a_{m,n,l} + \\
& J_y a_{m,n+1,l}^{\dagger} a_{m,n,l}+ 
K_z e^{-i\Phi_{m,n,l}} a_{m,n,l+1}^{\dagger} a_{m,n,l}+h.c. \left.\right). \nonumber
\end{align}
Here, $a_{m,n,l}^{\dagger}$ ($a_{m,n,l}$) is the creation (annihilation) operator on the site $(m,n,l)$,
and $\Phi_{m,n,l} = \delta{\bf k} \cdot {\bf R}_{m,n,l} = m \Phi_x+n \Phi_y+l \Phi_z$ 
are the nontrivial hopping phases, dependent on the positions ${\bf R}_{m,n,l}$. 
An inspection of Eq.~(\ref{Weyl H}) reveals that a wealth of energy dispersion 
relations can be achieved by manipulating the directions of Raman lasers 
$\delta{\bf k}$. Next, we choose the directions of the Raman lasers 
such that $(\Phi_x,\Phi_y,\Phi_z) = \pi (1,1,2)$, i.e.
$\Phi_{m,n,l} =(m+n)\pi$ (modulo $2\pi$). 
This is schematically illustrated in Fig.~\ref{lattice}(b). 
It should be noted that a seemingly equivalent choice  
$(\Phi_x,\Phi_y,\Phi_z) = \pi (1,1,0)$, will not be operational, 
because a {\em nonvanishing} momentum transfer in the tilt direction 
is necessary for the resonant tunneling to be 
restored~\cite{Miyake2013,Aidelsburger2013,MiyakePhD2013}.

A sketch of the 3D lattice obtained with such a choice of 
phases is illustrated in Fig. \ref{lattice}. 
It can be thought of as an alternating stack of two types of 2D lattices, 
parallel to the $xz$ plane, which are illustrated in \ref{lattice}(c) and \ref{lattice}(d); 
hopping between these planes is regular (along $y$). 
The 3D lattice has two sublattices (A-B). 
Another view is stacking of 2D lattices described by the Harper Hamiltonian $H_{\alpha=1/2}$
[Fig. \ref{lattice}(a)], such that the hopping along $z$ has phases $0$ or $\pi$, 
for $m+n$ even or odd, respectively. 
This breaks the inversion symmetry, and under application of Bloch's theorem, 
\begin{equation}
H({\bf k})=
-2 \{
J_y \cos(k_y a)\sigma_x + 
K_x \sin(k_x a)\sigma_y -
K_z \cos(k_z a)\sigma_z
\}.
\label{HW}
\end{equation}
Mathematically, the chosen phase engineering along $z$ has replaced the identity matrix 
in $H_{LN}$ with the Pauli matrix $\sigma_z$.

\begin{figure}
\centerline{
\mbox{\includegraphics[width=0.35\textwidth]{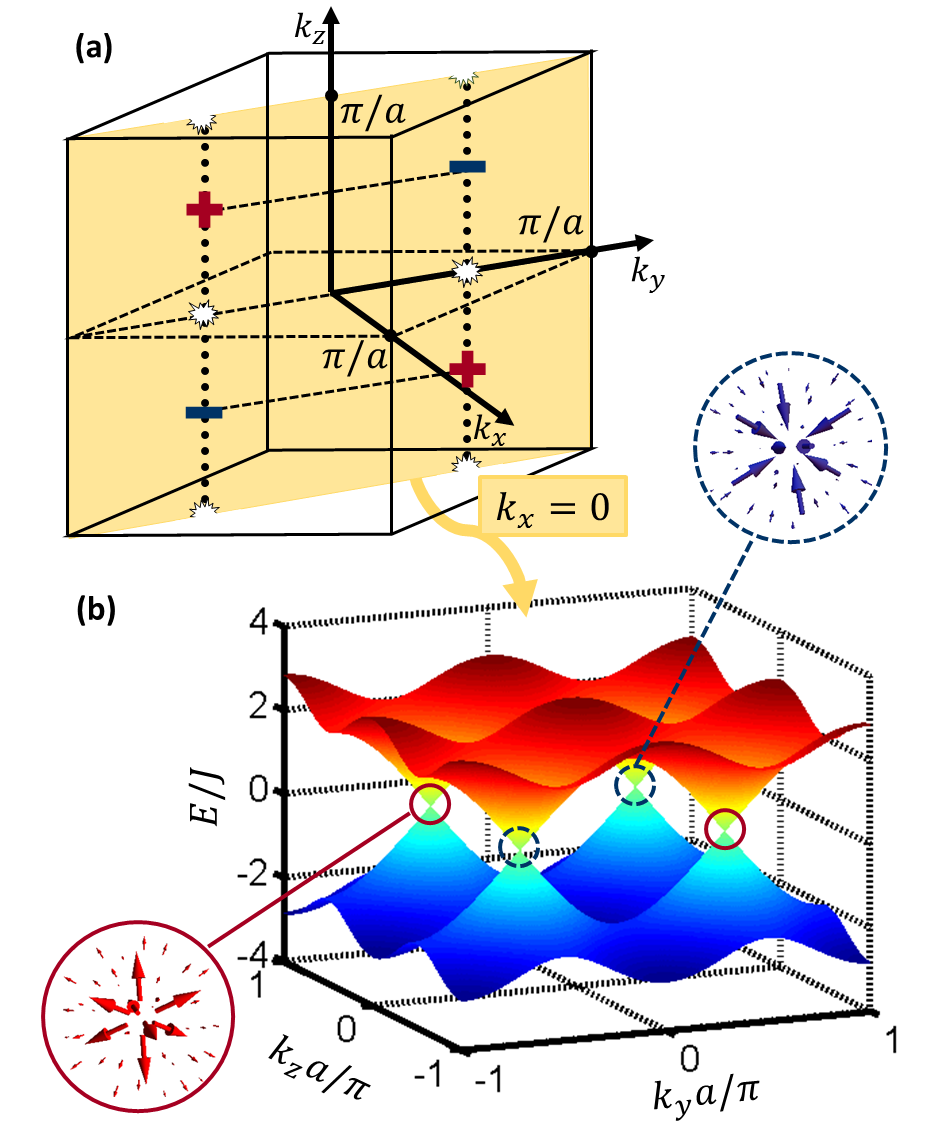}}
}
\caption{(color online) 
Sketch of the first Brillouin zone of the lattice depicted in Fig. \ref{lattice}, 
energy spectrum and Weyl points. 
(a) The positions of the Weyl points in the Brillouin zone and their chiralities are 
indicated with $+$ and $-$ signs. If a tunable A-B sublattice energy offset is 
introduced, Weyl points move along dotted lines, and can annihilate 
at points denoted with stars (see text). 
(b) Energy spectrum in the $k_x=0$ plane [shaded plane in (a)], 
shows linear dispersion in the proximity of the Weyl points. 
The insets show the Berry curvature of two Weyl points, demonstrating that 
they are synthetic magnetic monopoles in momentum space. 
}
\label{points}
\end{figure}

The energy spectrum of the Hamiltonian has two bands, 
\begin{equation}
E({\bf k})=\pm 2 \sqrt{K_x^2\sin^2(k_x a)+J_y^2\cos^2(k_y a)+K_z^2\cos^2(k_z a)}, 
\end{equation}
which touch at four Weyl points within the first Brillouin zone at $(k_x,k_y,k_z)=(0, \pm\pi/2a, \pm \pi/2a)$. 
Figure \ref{points} depicts the energy spectra in the first Brillouin zone, 
the Weyl points, and their chiralities. 
The dispersions around Weyl points are locally linear and described by the anisotropic Weyl Hamiltonian
$H_W({\bf q}) = \sum_{i,j} q_{i} \nu_{ij} \sigma_{j}$~\cite{HosurReview},
where ${\bf q} = {\bf k} - {\bf k}_W$ is the displacement vector from the Weyl point (located at ${\bf k}_W$) in momentum space. 
Here $[v_{ij}]$ is a $3\times 3$ matrix, with elements $v_{xy}=-2K_xa$, $v_{yx}=\pm 2J_ya$, 
$v_{zz}=\pm 2K_za$, and zero otherwise.
The topological nature of the system is reflected in the possibility to assign (positive and negative) chirality, defined as 
$\kappa=\mbox{sign}(\det[v_{ij}])$, to the Weyl points~\cite{Lu2013}.

Weyl points are monopoles of the synthetic magnetic field in momentum space. 
In order to verify this property of our energy nodes, we have calculated the gauge field, i.e. Berry connection
${\bf A}({\bf k})=i\langle u({\bf k}) |\nabla_{{\bf k}}| u({\bf k})\rangle$, and the 
synthetic magnetic field, i.e. Berry curvature ${\bf B}=\nabla_{{\bf k}} \times {\bf A({\bf k})}$. 
The obtained Berry curvature is depicted in the insets of Fig. \ref{points}, clearly demonstrating that what
we have proposed is a construction of topological synthetic magnetic monopoles in momentum space 
of a 3D optical lattice.

These monopoles are robust to any perturbation 
which adds a $\sigma_i$ term ($i = {x,y,z}$) to the Hamiltonian. 
The only way for Weyl points to disappear is when two of them with 
opposite chirality annihilate. This topologically protected nature of 
Weyl points can be probed in the proposed setup by adding a tunable A-B 
sublattice energy offset in the same fashion as in Ref. \cite{Jotzu2014}, such that 
the on-site energy at sites with $m+n$ odd (even) is $\epsilon$ ($-\epsilon$). 
This adds an $\epsilon\sigma_z$ term to the Hamiltonian in Eq. (\ref{HW}), 
and shifts the Weyl points parallel to the $z$-axis by tuning $\epsilon$, 
as illustrated in Fig.~\ref{points}(a). 
By making this term large enough ($\epsilon = \pm 2K_z$), one can drive the 
annihilation of the Weyl points pairwise either at 
$(k_x=0,k_y=\pm \pi/2a,0)$ for $\epsilon = -2K_z$, or at the edge of the 
Brillouin zone for $\epsilon = 2K_z$, and open up a gap in the system.

Now that we have identified the scheme which creates the Weyl Hamiltonian, 
we propose schemes for their experimental detection
which are applicable for both ultracold bosons and fermions. 
In order to verify that we have points at which the two bands touch 
in the 3D Brillouin zone, one can accelerate the initially 
prepared ultracold atomic cloud from the ground state position in 
momentum space towards the Weyl point using a constant force, 
and observe the crossover to the second band which can be revealed 
by time-of-flight measurements. 
By pushing the cloud in directions which would 'miss' the Weyl point, 
Bloch oscillations would be observed within the lowest band. 
Such a scheme was recently used to detect Dirac points in a honeycomb 
optical lattice~\cite{Taruell2012}, and also to probe the topological phase 
transition in the Haldane model~\cite{Jotzu2014}. 
Two points are worth emphasizing here. 
First, Weyl points are robust and would not be destroyed by an additional small force~\cite{AshwinReview,Lu2014}. 
Second, the trajectory of the gas being pushed would not be deflected 
in our lattice, because we have a time-reversal symmetric Hamiltonian.

The second scheme to observe the Weyl points is Bragg spectroscopy~\cite{Ernst2010}. 
By using an additional pair of Raman lasers, i.e., a two-photon Raman transition, 
one can couple states of the Hamiltonian (\ref{HW}) with a given energy and momentum difference, 
and induce excitations from the lower band to the upper band to probe the 
band-structure~\cite{Ernst2010}. 
This scheme would reveal the existence of Weyl points with very high resolution 
as it would not change the internal atomic state, and therefore not be 
sensitive to Zeeman shifts.

\begin{figure}
\centerline{
\mbox{\includegraphics[width=0.45\textwidth]{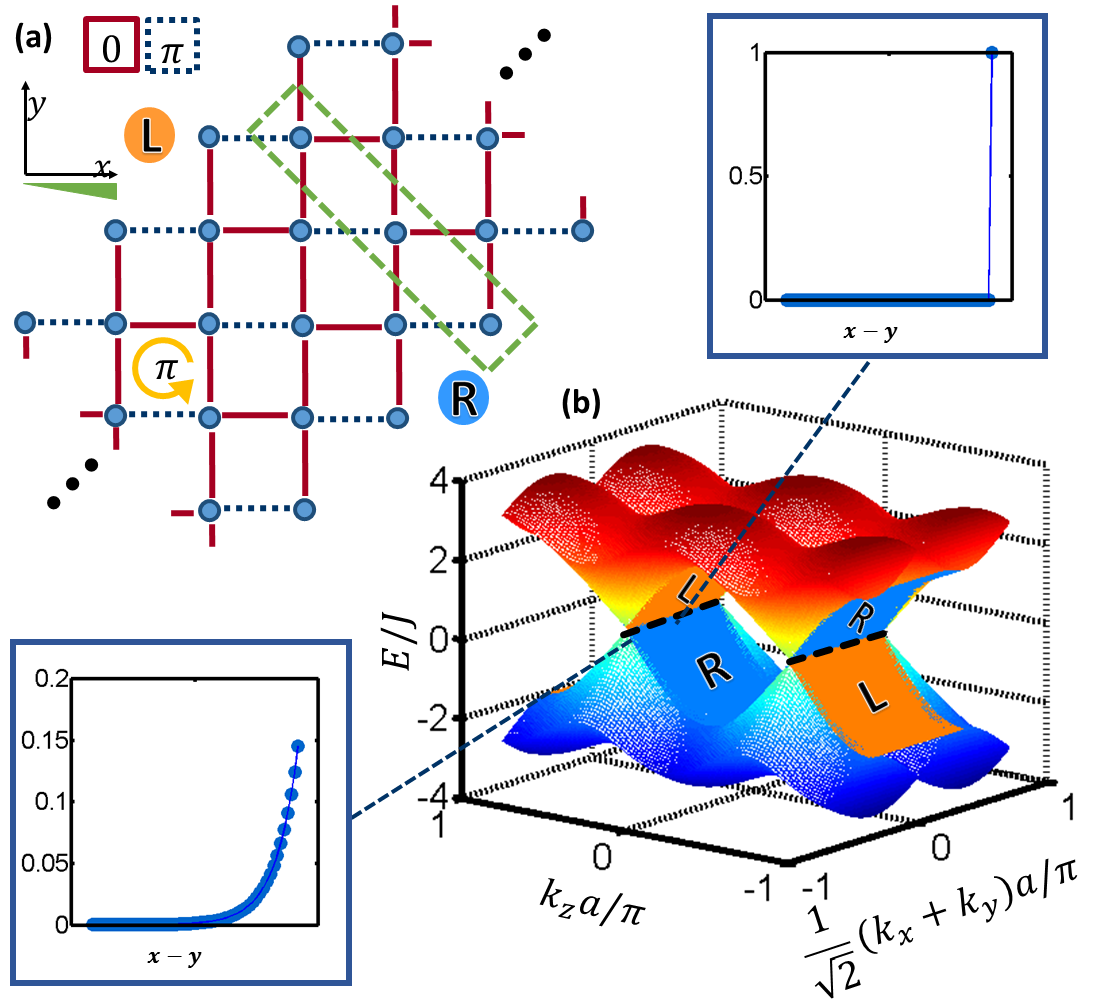}}
}
\caption{(color online) 
Surface states and Weyl points. 
(a) A slab of finite-width is cut from the 3D lattice along planes orthogonal 
to the $\hat{x}-\hat{y}$ direction; cross section in the $xy$ plane is sketched. 
The two sides of the slab are indicated with letters $L$ and $R$.
(b) Energy spectrum of the slab. The two dispersion sheets of surface states 
(corresponding to the two surfaces of the slab) are denoted with $R$ (blue) and $L$ (orange). 
The intersections of the two sheets are Fermi arcs (denoted with dashed lines).
The arcs connect Weyl points of opposite chiralities.
The insets show examples of the profile of the Fermi arc surface 
states (absolute value squared) across the slab, as indicated 
by the green dashed box in (a).
}
\label{fermiarcs}
\end{figure}

The proposed methods are applicable for both bosons and fermions. 
Here we discussed atoms in a single spin state, however, 
a mixture of spin states provides another degree 
of freedom to explore new phenomena, e.g., see ~\cite{CK2013}. 
By using single spin fermions, 
the Weyl semimetal phase could be achieved by adjusting the Fermi level 
to the energy of the Weyl points, that is, by properly 
adjusting the particle density.

Weyl semimetals imply the existence of intriguing topological surface 
states that come in the form of 'Fermi arcs' in momentum space~\cite{AshwinReview}. 
Topological effects such as Berry curvature have been experimentally observed in 
ultracold atomic systems~\cite{Aidelsburger2014,Jotzu2014}. 
However, surface states are difficult to detect with light scattering methods 
because one has to distinguish them from the bulk signal (e.g., see 
~\cite{Goldman2013} and references therein). 
Nevertheless, it is illustrative to show Fermi arcs and surface states in our model. 
In Fig. \ref{fermiarcs}(a) we take a slab of our lattice cut 
orthogonally to the $\hat x-\hat y$ direction (infinite along the $\hat z$
and $\hat x+\hat y$ directions), and in Fig. \ref{fermiarcs}(b) we plot 
the energy spectrum of this slab. 
The Weyl points of the infinite 3D lattice are now 
connected with 'Fermi arcs' in momentum space (shown with dashed lines). 
The states on the arcs are surface states~\cite{AshwinReview}, 
as can be seen from the inset in Fig. \ref{fermiarcs}(b)
(only states from one of the surfaces are shown). 
Surface states closer to the Weyl points spread more into the 
bulk than those in the center of the arcs. 
The Fermi arcs belong to two energy dispersion sheets of surface states, 
each one corresponding to one of the slab surfaces. 
The two sheets are located adjacent 
to the energy dispersion of bulk states~\cite{AshwinReview};
one sheet is on the bottom (the other is on the top) 
of the upper (lower, respectively) band. 
These two sheets intersect at Fermi arcs.

In conclusion, we have predicted the existence of Weyl points 
in the band structure of 3D optical lattices with 
phase-engineered hopping. Weyl points and 
chiral Weyl fermions have never been observed in nature. 
Experimental realization of Weyl points 
would enable exploring exotic topological states of matter, such as Weyl 
semimetals and associated topological surface states - 'Fermi arcs'~\cite{AshwinReview}. 
Here we pointed out that Weyl points, and all of the 
exciting phenomena that they include, could be experimentally addressed 
in 3D optical lattices with laser-assisted tunneling, 
in the experimental setup that was recently employed to obtain the 
Harper Hamiltonian~\cite{Miyake2013, Aidelsburger2013}.
Without phase engineered hopping, the creation of Weyl points would be more demanding,
possible only for a reduced number of space groups and 
points of symmetry in the Brillouin zone~\cite{Manes2012}.
An interesting venue would be to include interactions between the atoms, 
which can fundamentally change the system's behavior
(for an example, consider the interaction induced phase transition to a topological 
insulator in a fermionic 2D optical lattice~\cite{Sun_Sarma2012}).

This work was supported by the Unity through Knowledge Fund (UKF Grant No. 5/13),
the NSF through the Center for Ultracold Atoms, by NSF Award No. PHY-0969731, and an AFOSR MURI, 
and in part by the U. S. Army Research Laboratory and 
the U. S. Army Research Office through the Institute for Soldier Nanotechnologies, 
under contract number W911NF-13-D-0001.
We are grateful to Cody Burton, Woo Chang Chung, Liang Fu, 
John D. Joannopoulos, Mario Novak, Juraj Radi\'{c}, 
and Ashvin Vishwanath for useful discussions.


\end{document}